# Predicting Organic-Inorganic Halide Perovskite Photovoltaic Performance from Optical Properties of Constituent Films through Machine Learning

Ruiqi Zhang[1,2], Brandon Motes[5], Shaun Tan[3], Yongli Lu[3], Meng-Chen Shih[3], Yilun Hao[4], Karen Yang[1], Shreyas Srinivasan[3], Moungi G. Bawendi[3], Vladimir Bulović[1,2,*]

1. Department of Electrical Engineering and Computer Science, Massachusetts Institute of Technology, Cambridge, MA, 02139, US

2. Research Laboratory of Electronics (RLE), Massachusetts Institute of Technology, Cambridge, MA, 02139, US

3. Department of Chemistry, Massachusetts Institute of Technology, Cambridge, MA, 02139, US

4. Department of Aeronautics and Astronautics, Massachusetts Institute of Technology, Cambridge, MA, 02139, US

5. Optigon, Inc., Somerville, MA, 02143, US

\* Corresponding Author

**Abstract**

We demonstrate a machine learning (ML) approach that accurately predicts the current-voltage behavior of 3D/2D-structured $(FAMA)Pb(IBr)_3/OABr$ hybrid organic-inorganic halide perovskite (HOIP) solar cells under AM1.5 illumination. Our neural network algorithm is trained on measured responses from several hundred HOIP solar cells, using three simple optical measurements of constituent HOIP films as input: optical transmission spectrum, spectrally-resolved photoluminescence, and time-resolved photoluminescence, from which we predict the open-circuit voltage (Voc), short-circuit current (Jsc), and fill factors (FF) values of solar cells that contain the HOIP active layers. Determined average prediction accuracies for 95 % of the predicted Voc, Jsc, and FF values are 91%, 94% and 89%, respectively, with $R^2$ coefficients of determination of 0.47, 0.77, and 0.58, respectively. Quantifying the connection between ML predictions and physical parameters extracted from the measured HOIP films optical properties, allows us to identify the most significant parameters influencing the prediction results. With separate ML-classifying algorithms, we identify degraded solar cells using the same optical input data, achieving over 90% classification accuracy through support vector machine, cross entropy loss, and artificial neural network algorithms. To our knowledge, the demonstrated regression and classification work is the first to use ML to predict device photovoltaic properties solely from the optical properties of constituent materials.

**Introduction**

Hybrid organic-inorganic halide perovskite (HOIP) thin films have been extensively studied as promising candidates for photoactive layers in photovoltaic solar cells [1-5]. The high power conversion efficiency (PCE) of HOIP solar cells has been attributed to their low exciton binding energy, strong light absorption coefficients, long carrier diffusion lengths, and long carrier recombination lifetimes [6-8]. The large bandgap tunability underscores the potential in optimizing efficient HOIP light absorption across a wide wavelength spectrum, enhancing the light harvesting capabilities for improved solar cell PCE [9-11]. With the recent focus on 3D/2D structured perovskites, a maximum PCE of above 25% has been achieved through the improved management of charge carrier recombination, enhanced long-term operation stability, and incorporation of 3D/2D interfaces to form more defect-tolerant structure [12-14]. However, the efficiency of HOIP solar cells is predominantly influenced by the precursor recipes and fabrication processes, with a majority of state-of-the-art laboratory-scale cells lacking compatibility with large-area printable techniques such as slot-die coating or roll-to-roll coating [11, 15]. Thereby, with the increasing number of perovskite morphology designs, machine learning (ML) has emerged as a useful tool to help researchers analyze, simulate, and further forecast the experimental data [16-19]. Utilizing the computation advantages of processing large sets of data, researchers have previously applied ML to investigate HOIP device thermal stability [20], photoemission [21], capping layer choices [22], contact materials with thin film parameters [23, 24], long-term stabilities [25, 26], and to predict their optical behaviors [27]. Despite those dedicated efforts to advance ML-informed studies on perovskite solar cells, predicting the photovoltaic performances of the solar cell with a high regression accuracy remains a challenging endeavor [28-33], which we address in the present study.

To accurately model solar cell operation and account for the role of physical processes such as interfacial recombination between layers in the solar cell stack, band-bending, internal electric fields developed due to the charge extraction layers, and charge transport losses, physical-device-based models such as drift-diffusion [34, 35] and solar cell capacitance simulator (SCAPS) [36, 37] have been previously employed. They aim to determine the relevant simulation parameters from the measured experimental data fittings. However, the propensity of free variables in these models hinders their effectiveness as predictive tools for new device structures and solar cell performance [35-40].

Rather than using a detailed solar cell physical model, in this work we demonstrate a Blackbox ML approach that can accurately predict important features of merit of a solar cell, i.e., open-circuit voltage, short-circuit current, and fill factor. To generate the dataset for our ML model, we first made a set of several hundred HOIP solar cells and measured their current-voltage (J-V) characteristics under simulated AM1.5 illumination. We also measured their optical transmission spectrum (%T), spectrally-resolved photoluminescence (SrPL), and time-resolved photoluminescence (TrPL), as representative optical properties to include as our ML algorithm input. The ML algorithm is then trained to correlate these optical input data with the J-V characteristics, so that for a new HOIP device we can measure %T, SrPL, and TrPL and input them into the ML algorithm to predict the expected Voc, Jsc, and FF.

**Neural Network Training/Testing Dataset Generation and Algorithm Development**

Our experiments are staged as three common steps[31]: sample preparation & measurements, algorithms development, and model evaluation & optimization, shown in Fig. 1a.

Batches of solar cells are first fabricated with the structure of FTO/SnO$_2$ (KCl)/(FAPbI$_3$)$_{0.98}$(MAPbBr$_3$)$_{0.02}$/OABr/Spiro-OMeTAD(PMPIm)/Au (see Supplementary Information SI-Fig.S1). Following the previously reported method, SnO$_2$ layer is synthesized through chemical bath deposition where oxalic acid (OA) is used as linker molecules, along with a hydrogen peroxide (H$_2$O$_2$) and potassium chloride (KCl) treatment to minimize the interfacial non-radiative carrier recombination [3, 13]. These fabricated solar cells are then optically measured for their %T, SrPL, and TrPL via a high-throughput automated measurement system (SI-Fig.S2), and their J-V properties are separately recorded. Figs. 1b-d show %T, SrPL, TrPL, and J-V curves for a typical HOIP solar cell used in this study. This sample cell has a peak SrPL wavelength near $\lambda$ = 800 nm, which matches the bandgap energy of the perovskite active layer, and a PCE of 21.3%.

Each of the recorded %T, SrPL, and TrPL spectra have a data dimension of 2×N, where wavelength and intensity, or time and intensity, are 2 parameters measured for each of the N acquired data points. These 2-dimensional arrays become the inputs into the ML training algorithm. To optimize ML model convergence, the number of data points, N, in a training data set should be similar to the number of solar cells that will be measured [41-43]. In the present work, several hundred solar cells are fabricated and measured, and for each the three collected spectra are preprocessed to reduce the number of data points to 250 each. %T spectrum is collected over the wavelength range of $\lambda$ = 650 nm to 850 nm, and SrPL spectrum over the range of $\lambda$ = 700 nm to 900 nm. TrPL is similarly constrained over the lifetime span of 6 μs. We extract the rolling middle-point within 3-neighbor-point to extract 250 data points per spectrum, while keeping the x-axis (wavelength for %T and SrPL, time for TrPL) for each collected sample consistent across all samples (SI-Fig.S3). This allows us to eliminate all x-axis values of every spectrum as input into the ML algorithm, so that each spectrum turns into a one-dimensional array of 250 y-axis data points.

The artificial neural network (NN) algorithm we developed is illustrated in Fig. 2a. The three preprocessed input spectra, representing %T, SrPL, and TrPL, are initially vectorized into a 3×250 matrix. Subsequently, each vectorized column, undergoes a three-layer-neural-net to downgrade the learning matrix from 3×250 to 3×100 and further to 3×50, culminating in a 3×3 output. This process is followed by a concatenation layer that downgrades the 3×3 matrix into a 1×3 vector output representing the Voc, Jsc, and FF values. Throughout the forward training process, measured Voc, Jsc, and FF values are incorporated into the algorithm as ground truth to minimize the loss function through minimizing the L1-norm or L2-norm values.

To determine how the number of training samples might influence the final prediction, we first randomly select 10 experimentally measured samples as our testing dataset and then investigate how the NN algorithm performs in predicting these 10 testing samples as we vary the number of input training sample. As shown in SI-Fig.S4a, training the NN algorithm with less than 50 training samples leads to an averaged prediction error beyond 30%. However, with an increasing number of training samples, the averaged prediction error drops to less than 5% when 200 training samples are used.

To further optimize the NN prediction performance, we tuned two parameters in the algorithm: learning gradient descent steps, and number of training epochs. SI-Fig.S4 shows the stepwise training loss which tends to converge to a constant level after 7,500 training epochs when 0.0001 learning gradient descent rate is used. Therefore, we optimize the epoch number to be 15,000 to strike a balance between NN prediction accuracy and model training time.

**Neural Network Regression Predicting Results**

The initially fabricated and characterized 220 experimental devices exhibit a distribution of values for PCE, Voc, Jsc, and FF, as shown in SI-Fig.S5. These samples are then randomly separated into two groups: 210 training data and 10 testing data. After training the model based on L1-norm for 15,000 epochs (prediction architecture shown in SI-Fig.S6), a training error for all 220 training datasets is plotted in SI-Fig.S7, showing high fitting accuracy. Representative predictions of Voc, Jsc, and FF are shown in Fig. 2b, where the 10 testing data sets are plotted for three separate prediction attempts, showing a small prediction error. To further validate the neural network prediction outcomes, we investigate the influence of the datapoints distribution within the input dataset on the prediction error. In Fig. 2c, all measured data are grouped into 8 groups based on their PCE, while the three photovoltaic parameters (Voc, Jsc and FF) are also simultaneously categorized into 8 groups. We notice an improved average prediction accuracy for solar cells whose photovoltaic parameters align with the range of more extensive training datapoints, which is consistent with our finding in SI-Fig.S4a. The distribution demonstrates that with a larger portion of datapoints distributed in the input dataset, an enhancement in accuracy is achieved with %error reaching less than 5%. A repeated experiment with a smaller training input dataset is carried out to affirm the influence of data distribution within the input, as shown in SI-Fig.S8, reinforcing the consistency of this finding. Among the three predicted values, FF exhibits larger prediction fluctuations compared to the other two predicted outputs, which could be due to FF being affected not only by the device stack optical properties, but also by the solar cell series resistances and device fabrication artifacts that could affect the shunt resistance.

We carried out the "pressure test" on the algorithm by decreasing the amount of data in the training dataset while increasing the amount of data in the testing dataset. Typical ML algorithms expect the testing datasets to be less than 20% of the training dataset [18, 24, 27, 29]. Hence, in our initial training attempts, we opted for a training dataset of 210 data points and the remaining 10 data points for testing, resulting in a testing-to-training ratio of 10:210, or 5%. However, in the pressure test, we change the ratio of testing-to-training from 5% to 1,000%, i.e., 10-time-more data in the testing dataset as compared to the training dataset. As shown in Fig. 2d, the averaged %error increases as the test-to-train ratio increases, stabilizing at a level of 8% for Jsc, 10% for Voc and 20% for FF.

To further assess the prediction accuracy of the NN algorithm, we randomly select the 10-testing data from the dataset and predict their Voc, Jsc, and FF. We then repeat that procedure at least 10 additional times to make the algorithm predict over 100 tested data. Over 85% of all prediction errors fall under 5% of Voc, Jsc, and FF (as detailed in Fig. 3d). An overall averaged prediction accuracy of 97%, 98% and 95%, for Voc, Jsc and FF, respectively, with respective $R^2$ (coefficient of determination) values of 0.48, 0.06, 0.49, suggesting the robustness of the NN algorithm when predicting Voc and FF, shown in Fig. 3a-c and SI-Fig.S9. Note that the small spread in the training data values for Jsc leads to a relatively weak learning ability of this parameter indicated by small $R^2$ value. Hence to improve the prediction capability of Jsc, we further expand our dataset by adding to its data for degraded HOIP solar cells.

To generate a set of thermally degraded HOIP solar cell, fresh HOIP cells are heated at 80°, 100°, 120° and 150°C for 30 to 60 minutes. Degraded samples are then characterized for %T, SrPL, TrPL, Voc, Jsc, and FF following the same protocol shown in Fig. 1a. The expanded dataset comprises 368 samples, consisting of 220 initial cells and 148 degraded cells (SI-Fig.S10). The change in solar cell J-V characteristics with heating is shown in Fig. 3e and SI-Fig.S11. Changes in %T, SrPL, TrPL decay lifetime are plotted in SI-Fig.S12-16.

Using the same NN prediction algorithm on this larger training data set, with 0.0001 gradient learning steps and 20,000 epochs, prediction results are obtained and plotted in Fig. 3f-h and SI-Fig.S17. Notably, with the enlarged dataset that includes a broader distribution of the training datapoints, the Jsc $R^2$ value increases to 0.77, while $R^2$ for Voc and FF remain high at 0.47 and 0.58, respectively, when testing randomly selected 200 testing samples. Determined average prediction accuracies for 95 % of the predicted Voc, Jsc, and FF values are 91%, 94% and 89%, respectively.

To assess the algorithm's capacity for robust prediction with reduced input spectrum information, we also evaluate if the NN algorithm can be trained only by using one or two of the three collected optical spectra. Prediction errors of NN algorithms trained with two spectra (%T and SrPL, or %T and TrPL, or SrPL and TrPL) and one spectrum (only %T, or SrPL, or TrPL) are illustrated in Figs. 4b and c, respectively, with each of their learning $R^2$ value presented in Fig. 4a. As expected, results show a drop in $R^2$ value and an increase in %error as we decrease the set of training spectra from three to two to one. Thus, an enhancement in the number of training data sets largely boosts the model's ability to learn complex patterns and relationships within the data. Results of an identical investigation using the initial 220 sample dataset are shown in SI-Fig.S18, demonstrating the benefit of expanding the teaching data set.

**Connecting Neural Network Predictions to Physics Properties via Linear Regression**

Rather than using the 250 data points for each of the three optical spectra that are inputs to the NN algorithm, each of these input spectra can instead be described through a set of critical parameters that have a physical meaning. For example, the %T spectrum can be described by identifying five critical values, namely: (1) λ corresponding to the maximum %T; (2) Maximum %T intensity between λ=650-950 nm; (3) λ at which the transmission is 20% of the maximum; (4) λ at which the transmission is 80% of the maximum; (5) Average λ between 20% to 80% transmission. Similarly, additional 15 critical values can be used to describe SrPL and TrPL for the total of 20 physically-associated values that describe three collected optical spectra for each solar cell (see SI. Table 2 for the listing of parameters, SI-Fig.S19 and supplementary discussion 1 for the detailed data pre-processing procedure). We use these extracted 20 values for each of the measured solar cells as inputs into a separate prediction algorithm that is based on linear regression (LR, Fig. 5a). LR is then trained using the 20 extracted values from both the initial 220 samples (where 210 samples are used for training and 10 are used for testing) and the expanded set of 368 samples (where 348 are used for training and 20 for testing) to predict Voc, Jsc and FF. The LR training procedure is repeated multiple times with a random selection of training/testing samples. Examples of LR prediction results are shown in SI-Fig.S20.

As shown in Fig. 3a-d, the LR predicted output with 20 inputs parameters based on the initial dataset, is similar in accuracy to the NN algorithm. We note that $R^2$ for Jsc and FF is again small, which we again associate with a small spread in the training values. Repeating the same LR experiment with the expanded dataset shows improvement, with $R^2$ of 0.43, 0.54 and 0.47 for Voc, Jsc, and FF, respectively, as shown in Fig. 4a.

To identify which of the 20 parameters are most weighed in the LR training, we train the algorithm once again, but now using randomly-chosen 10 out of the 20 parameters. The prediction error of the LR algorithm with 10 input parameters is increased by up to 5% and $R^2$ value is decreased by 0.1 to 0.2 as compared to LR with 20 parameters. The experiment of randomly-choosing 10 parameters is then repeated $C_{20}^{10}$, i.e., 184,756 times, allowing us to generate sufficiently large data set of weights. By averaging the 184,756 predictions results, we determine the average normalized weights for each of the 20 parameters, together with the weights for original 20-input LR algorithm, which are plot in Fig. 5b. (SI-Fig.S21-22 show the detailed weights investigation). From the figure, we conclude that parameters 2, 3, 4, 7, 8, 13, 14, 15 are the dominating parameters in predicting output, while parameters 17, 18, 20 exhibit a moderate influence. Correspondingly, parameters 2, 3, 4 represent the steepness of the %T transition at the wavelength corresponding to the bandgap energy of the perovskite thin film. Parameters 7, 8, 13, 14, 15 describe the peak information of the SrPL curve. A steeper rise of the %T curve with wavelength and a narrower full-width at half-maximum (FWHM) of the SrPL curve would be consistent with a more crystalline (less amorphous) perovskite film in the HOIP solar cells, which might lead to a higher PCE. Therefore, we examined if the PCE of the measured devices is corelated to their linearly fitted %T spectrum slope and SrPL FWHM, as shown in SI-Fig.S23. We find that a smaller SrPL FWHM leads to a higher PCE, while we do not see a significant dependance of PCE on the slope of %T.

We notice that TrPL data is less weighed in the LR predictions. To double-check this finding, we train the LR algorithm again, but this time using only two or one input spectra (Fig. 5c-d). Again, we find that highest $R^2$ values are obtained when %T and SrPL are used as the pair of input spectra, with lower $R^2$ values obtained when TrPL spectra are one of the two spectral inputs (see also SI-Fig.S24). In general, limiting the number of input spectra dampens the LR learning ability as indicated by the increase in prediction error and a drop in $R^2$ value, which is further validated in Fig. 5d (and SI-Fig.S24 d-f, SI-Fig.S25-27 for small dataset) where we show the LR algorithm results when trained by using only one input spectrum. An additional investigation on potential nonlinear relationship prediction (polynomial regression, PR) is discussed in supplementary discussion 2 and SI-Fig.S28-29, extending the consistency of our conclusions.

**Enhanced Classifier Algorithms**

To perfect the solar cell performance, it is important to optimize the operation of the well-performing cells. Therefore, we also developed classifier algorithms that can separate the poorly-performing cells from the well-performing ones. The aim of our classifier algorithms is to discretize the 368 training data sets into three groups, based on their photovoltaic performance: good cells (Voc >900 mV, Jsc >20 mA/cm$^2$, and FF >60%), OK cells (Voc between 700-900 mV, Jsc between 17-20 mA/cm$^2$, and FF between 45-60%) and bad cells (Voc <700 mV, Jsc <17 mA/cm$^2$, and FF <45%).

Our first classifier algorithm is based on the cross-entropy-loss (CEL) classification, that is suitable for multiclass classification problems, and which assigns a probability to each of the predicted classes. Fig. 6a shows that for 40 example predictions, most of the predicted categories are assigned a 100% probability, but in few cases the classification algorithm is less certain, and it assigns a percentage probability for two predicted categories. One limitation of the CEL classification is that the algorithm can be sensitive to outliers and noisy data, as it heavily penalizes misclassifications. Hence, we also developed the support vector machines (SVM) and Artificial Neural Network classifier (ANN-classifier) algorithms as alternatives. SVM is effective in high-dimensional spaces, making it suitable for tasks that could involve complex relationships. The ANN-classifier algorithm is coded similarly to the NN regression algorithm, however, substituting the original output Voc, Jsc, and FF values into discrete values of 0, 1, 2, representing classification categories of bad, OK, and good, respectively. Sample classification predictions of SVM and ANN-classifier are plotted in Fig. 6b, c. Fig. 6d illustrates the statistical distribution of classification predictions based on the comparison of ANN-classifier, CEL, and SVM algorithms. Similar accuracy is observed for all three classifiers, with an average accuracy above 80%.

To further correlate the three photovoltaic parameters with the overall cell performances, for each of the Voc, Jsc, and FF parameters we assign a score of 0, 1 and 2 to the bad, OK and good performance cells. An overall cell performance score is then the product of the three scores of that cell. A cell performance score of 0 is produced when either Voc, Jsc, or FF presents 0. A cell performance score of 8 is produced when Voc, Jsc, and FF are all well-performing. The predicted classification results are then also calculated following this rubric and compared with the ground truth values. Shown in Fig. 6e, all three algorithms perform with more than 90% accuracy, showing that they can act as both reliable regression algorithms and classification algorithms.

Through this work, we demonstrated ML algorithms that can accurately forecast photovoltaic properties of HOIP solar cells from the optical measurements of their constituent photoactive films. Our work shows the need for a comprehensive and extensive training dataset, the development of robust and efficient algorithms, the interpretation of linkages between machine learning models and physical material properties, and the extension of regression algorithms into classification algorithms. The work showcases the potential of ML in optimization of the design of future HOIP solar cells, with the aim of accelerating the manufacturing process in the solar industry.

**Methods**

**Materials**

All materials in this study were used without further purification, including lead iodide ($PbI_2$, Sigma-Aldrich), lead bromide ($PbBr_2$, Sigma-Aldrich), dimethylformamide (DMF, Sigma-Aldrich), dimethyl sulfoxide (DMSO, Sigma-Aldrich), methylammonium chloride (MACl, Greatcell Solar), formamidinium iodide (FAI, Greatcell Solar), methylammonium bromide (MABr, Greatcell Solar), n-octylammonium bromide (OABr, Greatcell Solar), DI water (Sigma-Aldrich), hydrochloric acid (HCl, 37 wt. % in water, Sigma-Aldrich), oxalic acid (purified OA, 99.999% trace metals basis, Sigma-Aldrich), Urea (ACS reagent, 99.0-100.5%, Sigma-Aldrich), tin(II) chloride dihydrate ($SnCl_2 \cdot 2H_2O$, >99.99%, Sigma-Aldrich), thioglycolic acid (TGA, 98%, Sigma-Aldrich), hydrogen peroxide solution ($H_2O_2$, 20 wt%, Sigma-Aldrich), diethyl ether (Sigma-Aldrich), chloroform (Sigma-Aldrich), chlorobenzene (Sigma-Aldrich), triphenylphosphine oxide (TPPO, Sigma-Aldrich), isopropyl alcohol (IPA, Sigma-Aldrich), N-Propyl-3-methylpyridinium bis(trifluoromethylsulfonyl)imide (PMPIm, Strem Catalog), and 2,2',7,7'-tetrakis(N,N-di-p-methoxyphenylamino)-9,9'-spirobifluorene (Spiro-OMeTAD, LT-S922H, Lumtec). Gold pellets used in the deposition of electrode materials were purchased from Kurt J. Lesker.

**Device Fabrication and Characterization**

Preparation of solutions from which HOIP solar cells were fabricated followed the method described in Lu, *et al*. [3].

SEM images were taken at MIT.nano using a Zeiss Gemini 450 SEM. J-V measurements were carried out using a Keithley 2400 source meter in the dark and while the devices were illuminated with simulated AM 1.5 irradiation generated by a xenon-lamp-based solar simulator (Oriel Sol3A, 94043A). To avoid small-area photovoltaic device edge effects during J-V measurements, we placed a shadow mask over the solar cells during illumination that exposed only the active cell area to the incident light.

The %T, SrPL and TrPL were measured in collaboration with *Optigon, Inc*. The measurement system is composed of a motion rail, a batch sample holder, a multiple laser head and an optical detector (see Fig. 1a and SI-Fig.S2). The motion rail transferred 5 samples at a time into the measurement module where for each sample %T, SrPL and TrPL is measured sequentially in the span of 3 seconds.

SrPL measurement was carried out using a continuous-wave (CW) laser with a laser power of 810 nW and an integration period of 500 ms. TrPL spectra were measured with a pulsed laser with a pulse width of less than 0.1 ns, laser power of 19 pJ/cm$^2$ per pulse, frequency of 200 kHz, integration period of 500 ms, and bin resolution of 10.24 ns.

The %T was acquired by shining a broadband white LED source over an integration period of 200 ms with an incident power of 50 nW. All three characterizations were measured in a dark box at room temperature.

**Algorithm Developments**

All machine learning algorithms are developed using python. Experimental data is developed with Optigon, Inc. packages *optigon.dataframe* and *optigon.mockdata.generation*. Neural network algorithm was developed based on *pytorch* package. Linear/polynomial regression algorithm utilized *sklearn.linear_model* package. Input curves were fitted and preprocessed through *scipy.optimize* and *sklearn.preprocessing*. Classifier algorithms (ANN, CEL, and SVM) were formulated by incorporating a modified cross-entropy loss function derived from the pre-established NN algorithm and utilizing the *sklearn.SVM* package.

**Data availability**

Paper is currently under review. Selected codes will be available on GitHub later.

**Acknowledgements**

We thank Dr. Brain Anthony for discussions on the algorithm development. We thank Tamar Kadosh, Jeremiah Mwaura and Tong Dang from MIT Organic and Nanostructured Electronics Lab (ONE Lab) for their many discussions. We thank all colleagues from Optigon in acquiring the data. This work was supported by the US Department of Energy (DOE, grant number DE-EE0009512). We thank Timothy Siegler and Marie Mapes from DOE for their insightful comments during our project meetings. The microfabrication and characterization in this work was in part performed at MIT.nano, MIT Bawendi Lab (Dept. of Chemistry) and MIT ONE Lab (Dept. of EECS).

**Author Information**

Authors and Affiliations

--Department of Electrical Engineering and Computer Science, Massachusetts Institute of Technology, Cambridge, MA, 02139, USA

Ruiqi Zhang, Karen Yang, Vladimir Bulović

--Department of Chemistry, Massachusetts Institute of Technology, Cambridge, MA, 02139, USA

Shaun Tan, Yongli Lu, Meng-Chen Shih, Shreyas Srinivasan, Moungi G Bawendi

--Department of Aeronautics and Astronautics, Massachusetts Institute of Technology, Cambridge, MA, 02139, USA

Yilun Hao

--Research Laboratory of Electronics (RLE), Massachusetts Institute of Technology, Cambridge, MA, 02139, USA

Ruiqi Zhang, Karen Yang, Vladimir Bulović

-- Optigon Inc., Somerville, MA, 02143, US

Brandon Motes

Author Contributions

V.B. and R.Z. conceived the idea. R.Z., V.B., and M.B. developed and optimized the idea. Y.L., M.S., T.S., and R.Z. contributed to the device fabrications and aging under the guidance of M.B. R.Z. and B.M. carried out the %T, SrPL, and TrPL characterizations. R.Z. and K.Y. carried out the solar cell J-V-measurement. Y.L. carried out the SEM measurement under the guidance of M.B., R.Z., and Y.H. developed and debugged the algorithms. R.Z., and B.M. developed the measurement codes. R.Z. and S.S. contributed to the data fitting. R.Z. contributed to the schematics and photographs. All authors contributed to the discussion of data and commenting on the manuscript.

Corresponding author

Correspondence to Vladimir Bulović

**Competing interests**

NA

**Main Figure Captions**

**Fig. 1 | HOIP solar cell structure, device properties, and fast-acquisition optical measurement system. a**. Three main steps are illustrated from left to right: I. HOIP solar cell structure, II. Automated collection of the solar cell optical spectra, and III. ML algorithm development with spectral information as input. **b-d**. Example device properties obtained from a solar cell sample. The obtained properties are used as the training data for ML algorithms and as ground truth. **b, c**: % transmission (%T) with spectrally-resolved photoluminescence (SrPL), and time-resolved photoluminescence (TrPL) are used as the ML training input. **d**. J-V characterization generates the values of Voc, Jsc, and FF for each device, which are also the output values generated by the ML algorithm. The example measured solar cell has a Voc of 1.11 V, Jsc of 24.67 mA/cm$^2$, FF of 77.7% and a PCE = Voc×Jsc×FF of 21.29%.

**Fig. 2 | Neural Network algorithm for predicting perovskite photovoltaic properties. a**. Neural Network Algorithm workflow. Three preprocessed optical spectra of all training samples are first vectorized into matrix form, with each matrix column then entered as input into the neural network. Each spectrum is processed by three neural-net-layers that reduce each of their dimension from 250 points into 3 points, followed by a concatenation layer that further shrinks the total of 9 values (3 from each spectrum) into 3 output values, corresponding to Voc, Jsc, and FF. Training outputs after each epoch are then optimized by minimizing the loss function value from the ground truth to obtain the final converged prediction results. **b.** Example of 30 predicted Voc, Jsc and FF results (light gray) with the comparison of their real experimental value (bolded black), demonstrating a small error obtained for the testing dataset. Calculated percent error for each prediction is shown in the bottom in blue. **c.** Influence of the number of similar data points in a training data set on the prediction error in that data range. Averaged percent error for Voc, Jsc and FF are shown in black, blue, and red, respectively. **d.** Influence of the number of data points in the testing dataset vs. training dataset on prediction error. Average prediction error increases as fewer datasets are used to train the machine learning model. The x-axis shows the ratio of the testing data to the training data: 10:210 (5%), 20:200 (10%), 70:150 (50%), 110:110 (100%), 145:75 (200%), 165:55 (300%), 184:36 (500%), 200:20 (1000%).

**Fig. 3 | Training accuracy with initial (220 HOIP solar cell samples) and extended (368 samples) data set, and the degradation of the HOIP solar cell under heat treatment.** Parity plots of 225 ML predicted values (y-axis) based on NN (Red) vs. LR (blue) for **a.** Voc, **b.** Jsc, and **c.** FF compared with their ground truth experimentally measured values (x-axis). The predicted $R^2$ values corresponding to NN and LR algorithms are indicated in the legend of the plot. **d.** Comparison of the prediction result accuracy for neural network (full spectrum as input), linear regression (selected 20 and 10 parameters as input), and polynomial regression (selected 20 parameters as input). Statistical distribution is obtained from over 100 prediction runs. **e.** HOIP solar cell photovoltaic properties decay after thermally treating the cells at 80º, 100º, 120º and 150ºC for 60 minutes. A drop in Voc, Jsc and FF is observed (also See SI-Fig.S11). **f-h.** Parity plots of 200 random examples of measured Voc, Jsc, and FF ground truth values compared with their NN-predicted results that were trained on the extended dataset (368 solar cell samples). An enhanced $R^2$ is obtained for Jsc prediction result after extending the dataset, while $R^2$ values for FF and Voc remained similarly high.

**Fig. 4 | Comparison of coefficient of determination R² and %Error for Neural Network algorithms trained with different input data sets and the Linear Regression algorithm. a.** $R^2$ values for traditional Neural Network (NN_Orig), Linear Regression with 20 parameters (LR_20_Para), and compressed-input Neural Networks that are trained on combinations of two spectra or a single spectrum. A sharp decrease in $R^2$ value is observed when decreasing the number of spectral inputs into the algorithms. Embedded image is the statistical prediction %error result of LR-20-parameters for over 200 testing samples, related to the weights investigation in Fig. 5b-d. **b.** Comparison of NN prediction result errors for extended dataset, obtained when using two out of the three measured spectra for training: The two spectra are either %T & SrPL, or %T & TrPL, or SrPL & TrPL. A similar level of averaged % error but enhanced $R^2$ value (bottom blue numbers) are observed among these three training experiments. **c.** Comparison of NN prediction result errors, obtained when using only one out of the three measured spectra for training: either %T, or SrPL, or TrPL for enlarged dataset. A further decrease in $R^2$ value is observed for all three experiments, validating the importance of using all three spectra in training of a robust NN prediction model. Each training is done for more than 200 times to generate the statistical errors distribution.

**Fig. 5 | Regression algorithm as preparation for physical algorithm understanding. a.** Logistics of linear/polynomial regression algorithm for physical parameters analysis. From the raw data for % T, SrPL, and TrPL, each of which is a set of 250 data points for each device, we extract 20 parameters that correspond to these physical measurements. We use these 20 parameters for each device as the input into the linear regression ML algorithm. The simplified data set allows us to see which specific physical parameters are weighed the most in the ML algorithm. Again, the input parameters are vectorized into the fitting algorithm to obtain the predicted weights. These final weights are trained through the cost function feedback equation that have minimized the residue error. **b.** 20 parameter weights based on linear regression prediction algorithm for extended dataset. Solid bins represent the weights predicted directly from 20 parameters. Dash dot line represents the averaged weights after randomly selecting 10 parameters out of the 20 for 184,756 training times. Blue background area encompasses parameters generated from %T, yellow encompasses SrPL parameters, and green encompasses TrPL parameters. **c-d.** Prediction errors with $R^2$ value based on shrink-training-input LR algorithm to further investigate parameters weights. Weights corresponding to each measurement are shown in SI-Fig.S24. Each training is done for more than 100 times to generate the statistical errors distribution.

**Fig. 6 | Classifier Algorithms with category prediction accuracy. a**. 40 representative cross-entropy loss results in predicting three categories of solar cells: good, OK and bad. Grey level shadings, from white to black, represent the percent of certainty, from 0% to 100%, in the prediction. Red dots in each bin represents the ground truth category of each experimental value. Three columns from left to right represents Voc, Jsc, and FF, respectively. **b.** 100 representative SVM classification results with the comparison of their ground truth categories. **c.** 60 representative ANN classification results with the comparison of their ground truth categories. The ANN algorithm gives a decimal predicted value between -0.9 to 2.9. A rounding procedure is processed to get the nearest integer 0, 1, or 2, representing bad, OK and good categories. **d.** A classification accuracy comparison based on three models. A comparably better classification accuracy is achieved for ANN-classifier algorithm. A better classification accuracy is observed for predicting Jsc and FF values. **e.** Final HOIP solar cell photovoltaic performance category prediction pie-chart results based on the scoring rubric described in the text. Blue section represents a bad-cell that is predicted to be a good-cell (false positive), and vice versa for the pink section (false negative). A yellow region represents a correct cell performance classification prediction.

# Predicting Organic-Inorganic Halide Perovskite Photovoltaic Performance from Optical Properties of Constituent Films through Machine Learning

# Main Figures

Ruiqi Zhang[1,2], Brandon Motes[5], Shaun Tan[3], Yongli Lu[3], Meng-Chen Shih[3], Yilun Hao[4], Karen Yang[1], Shreyas Srinivasan[3], Moungi G. Bawendi[3], Vladimir Bulović[1,2,*]

1. Department of Electrical Engineering and Computer Science, Massachusetts Institute of Technology, Cambridge, MA, 02139, US

2. Research Laboratory of Electronics (RLE), Massachusetts Institute of Technology, Cambridge, MA, 02139, US

3. Department of Chemistry, Massachusetts Institute of Technology, Cambridge, MA, 02139, US

4. Department of Aeronautics and Astronautics, Massachusetts Institute of Technology, Cambridge, MA, 02139, US

5. Optigon, Inc., Somerville, MA, 02143, US

\*   Corresponding Author

# Fig.1 Device Properties and Measurement System

**a.**

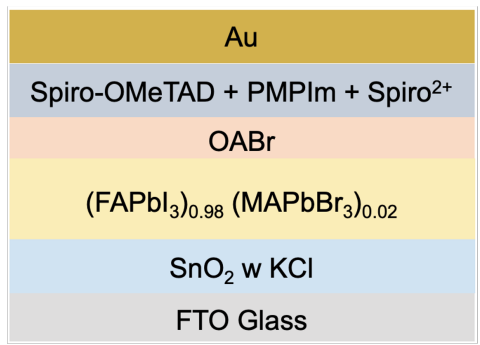

I. Sample Preparation

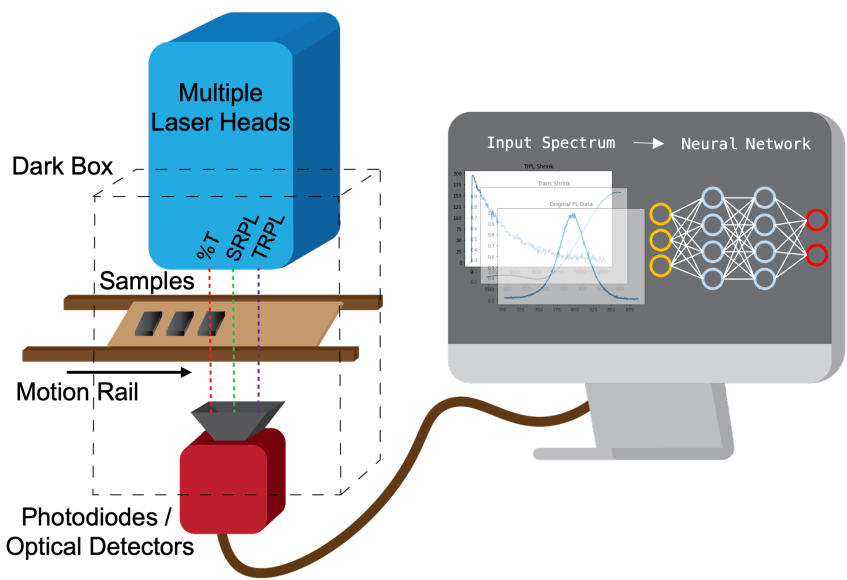

II. High-speed Measurement

III. Algorithm Developments

**b.**

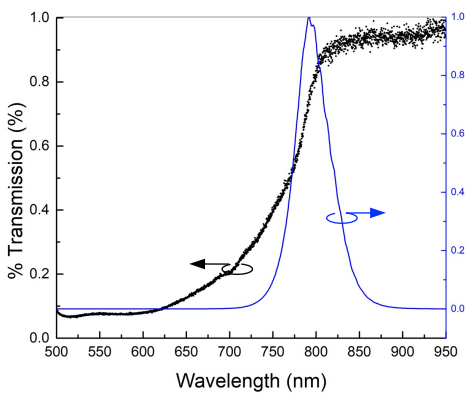

**c.**

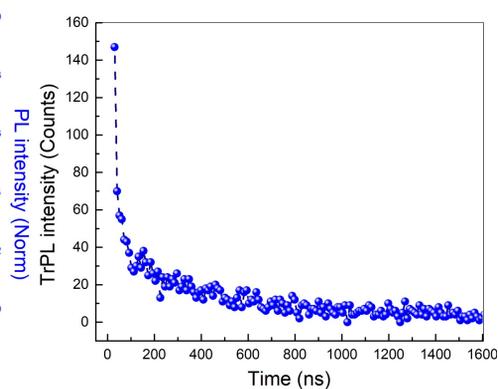

**d.**

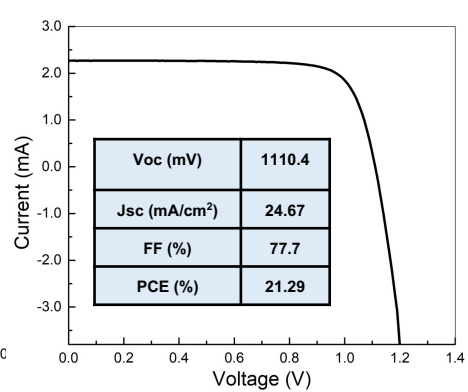

**Fig.2 Neural Network and Core Prediction Result**

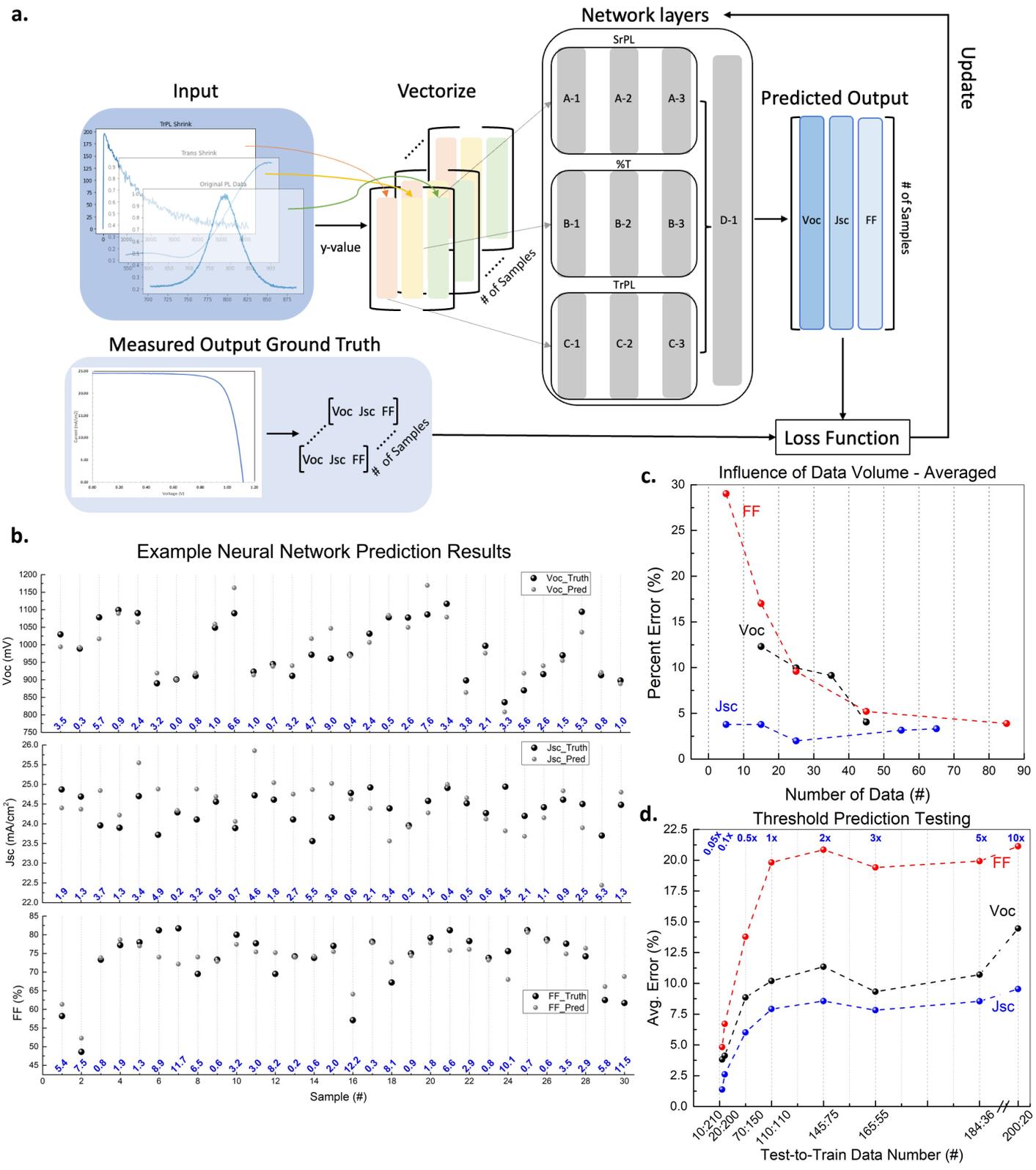

**Fig.3 Neural Network Accuracy Comparison and Enlarged Dataset**

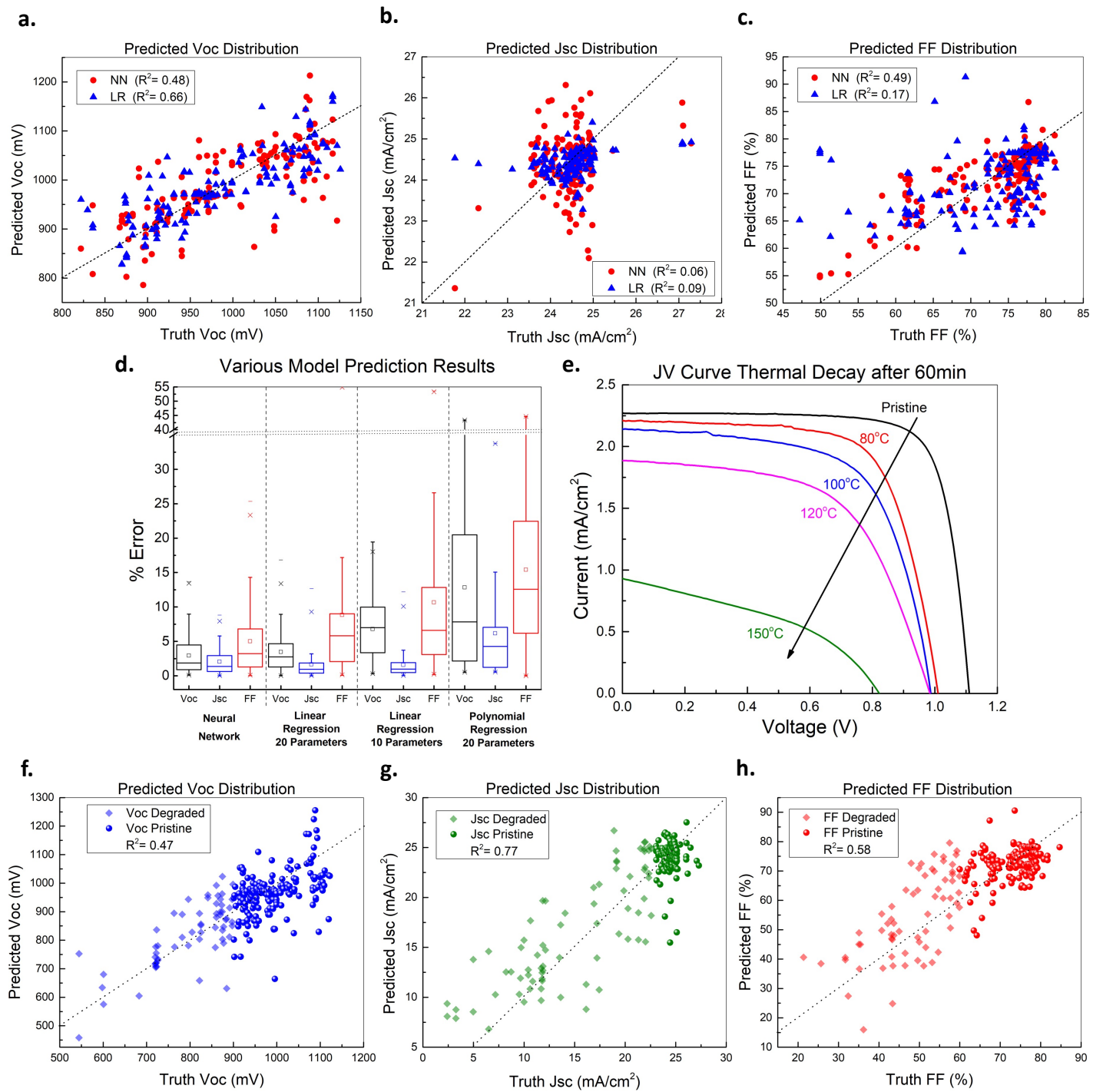

**Fig.4 Shrink NN Algorithm Prediction**

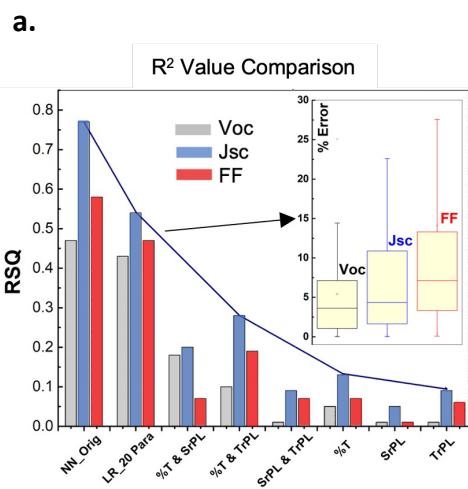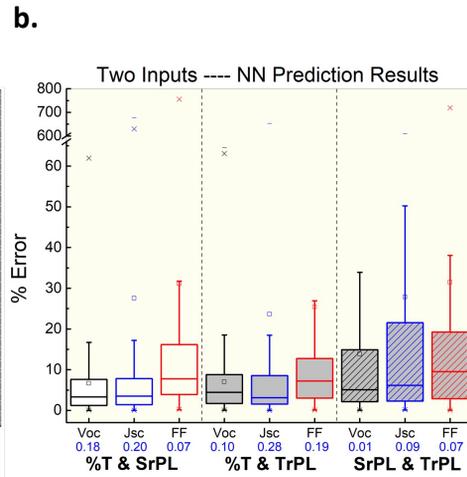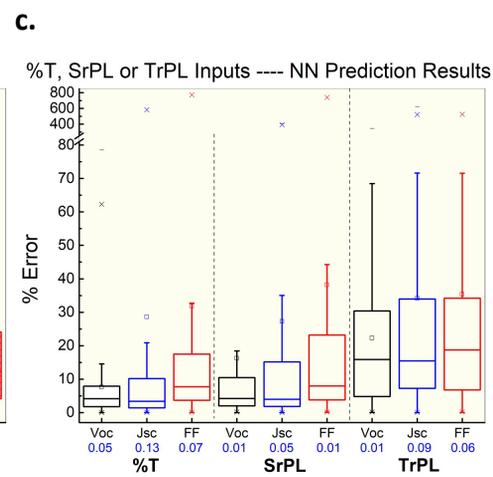

**Fig.5 Linear Regression Weights ---- Physical Understanding**

a.

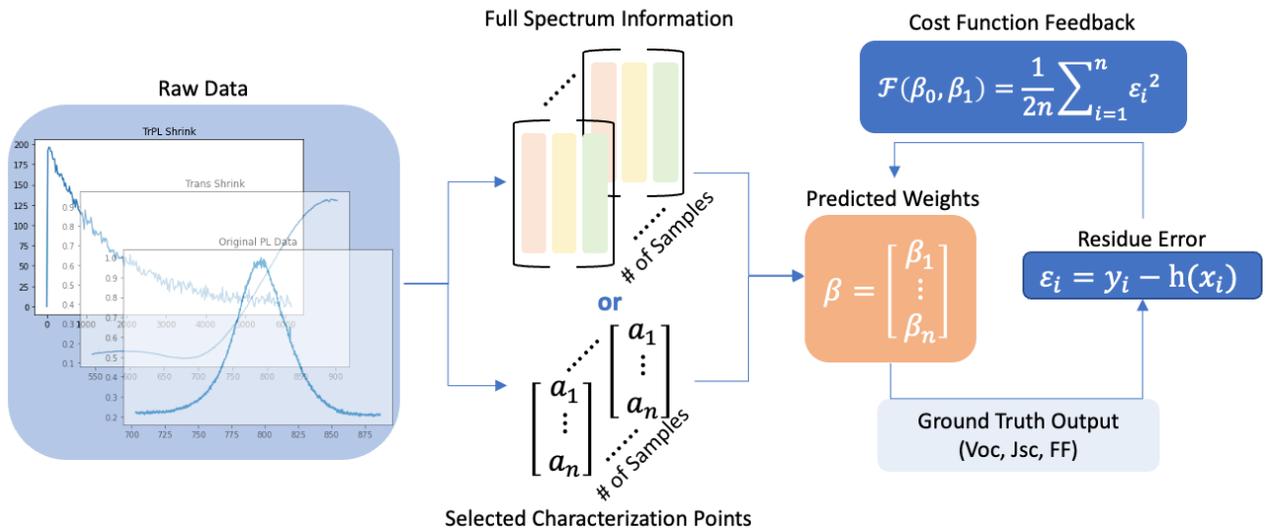

b.

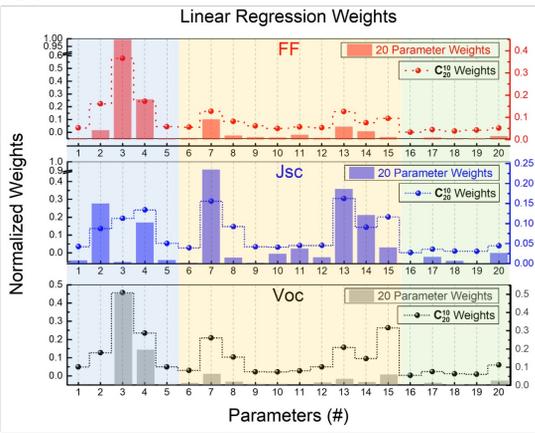

c.

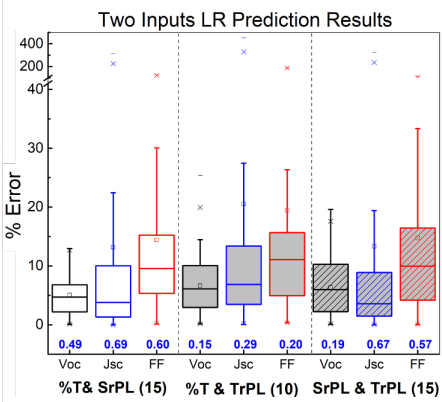

d.

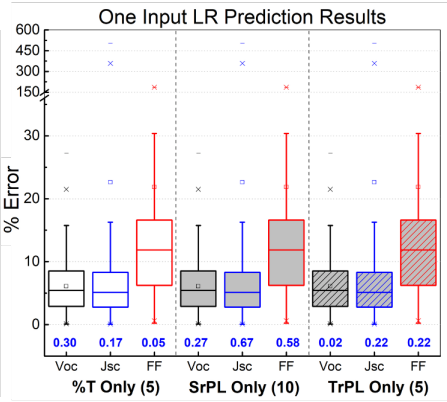

**Fig.6 Support Vector Machine and Classifier**

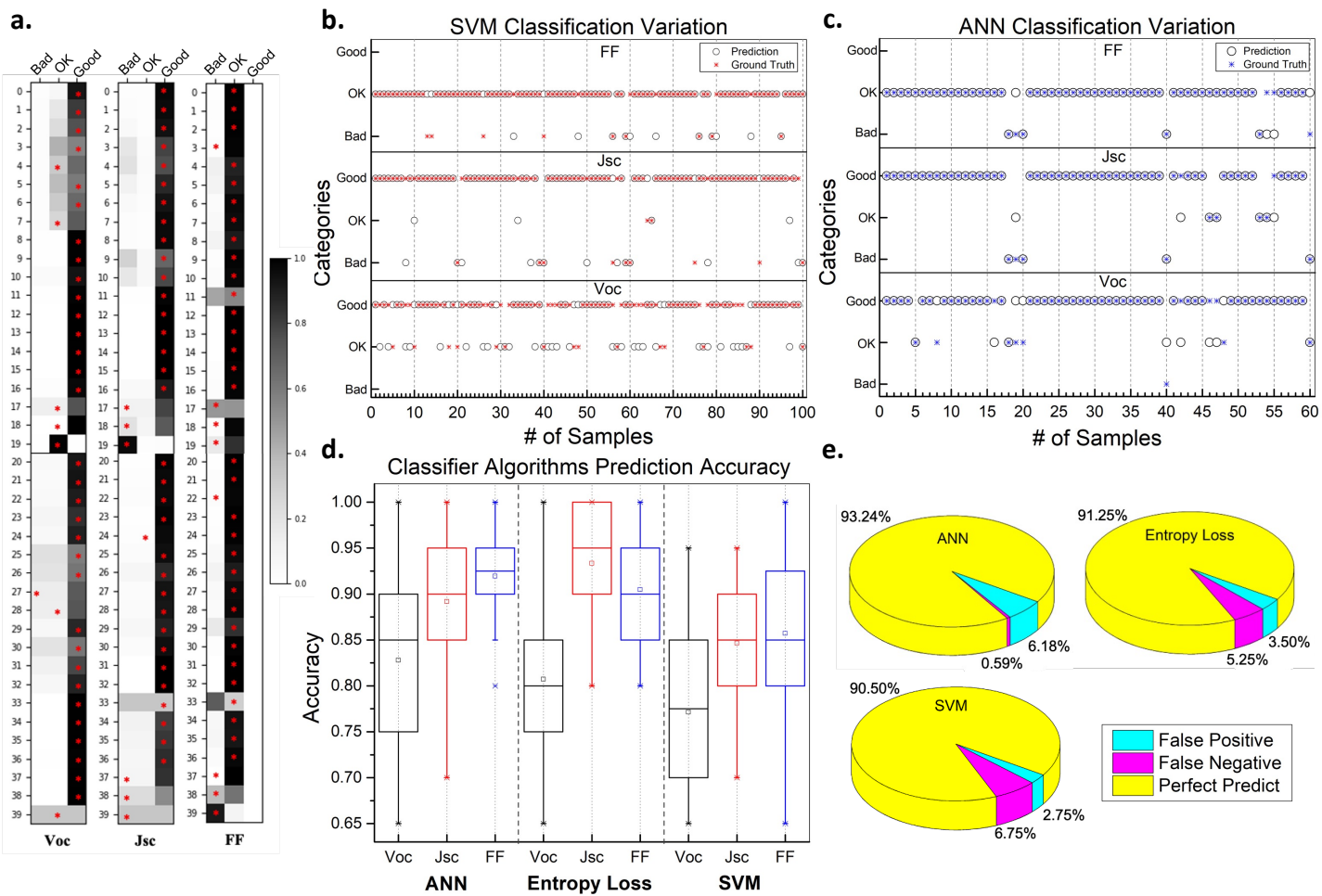